\documentclass{article}

\usepackage[utf8]{inputenc}
\usepackage[T1]{fontenc}
\usepackage[margin=1in]{geometry}
\usepackage{amsmath}
\usepackage{amssymb}
\usepackage[dvipsnames]{xcolor}
\usepackage[section]{placeins} % keep floats within section they are written
\usepackage{lineno}
\usepackage{graphicx}
\usepackage{parskip} 
\usepackage{xr}
\usepackage{caption}
\usepackage{subcaption}
\usepackage{float}
\usepackage{comment} %to access block comments
\usepackage{hyperref} %for url to github repo
\usepackage{textcomp, gensymb} %remove errors, degree command
\usepackage{rotating}

\usepackage[style=authoryear, sorting=nyt, sortcites=true, backend=biber, natbib=true, maxcitenames=2]{biblatex}
\DeclareDelimFormat[parencite]{nameyeardelim}{\addspace}
\usepackage{titling}
\predate{}
\postdate{}
\date{}

\providecommand{\keywords}[1]{\small \textbf{\textbf{Keywords --}} #1}

\newcommand{\CM}{\mathcal{C_M}}
  
\title{Tipping to Climate Action: Qualitative Insights from a Social-Climate Model with a Committed Minority}
\author{Sarah K. Wyse, Eric Foxall, Rebecca C. Tyson$^{\star}$}

\begin{document}

\maketitle
%\linenumbers

Department of Computer Science, Mathematics, Physics and Statistics, Irving K. Barber Faculty of Science, University of British Columbia Okanagan, SCI354, 1177 Research Road, Kelowna V1V 1V7, BC, Canada\\
$^{\star}$ Corresponding author. \\
\textit{Email addresses}: swyse@student.ubc.ca (S.K. Wyse), eric.foxall@ubc.ca (E. Foxall), rebecca.tyson@ubc.ca (R.C. Tyson).\\
\textit{ORCID IDs}: 0000-0002-5500-2711 (S.K. Wyse), 0000-0003-1610-6662 (E. Foxall), 0000-0002-7373-4473 (R.C. Tyson).

\begin{abstract}
    It is well-established that human activity is driving extreme weather patterns, and that these extreme events influence human behaviour. However, few models allow for human behaviours and the climate to dynamically interact. The models presented in this paper expand on previous work and serve as an initial framework to extend current models by using a dynamic social-climate feedback loop. First, we introduce a social model to determine the conditions under which a committed minority can overturn a pre-established social convention. Second, we modify an existing climate model to include climatic variability. Lastly, we formulate a social-climate feedback model to study the interplay between human behaviour and the climate. Our results demonstrate that the social-climate feedback loop may be important in accurately predicting future temperatures, in contrast to the standard approach where human behaviour is {\it a priori}. Additionally, we find that a committed minority plays a vital role in shifting public opinion towards climate action and that the time at which the social convention of climate inaction is overturned has a large impact on future temperatures. 
\end{abstract}


\section{Introduction} \label{Sect:Introduction}
%--------------------------------------------------------------------------------
%CLIMATE CHANGE---------------------------------------------------------------------
At the current warming of 1.55 degrees Celsius \citep{WMO2025}, we are experiencing extreme heat waves, flooding, fires, and storms \citep{IPCC_data}. Studies suggest that the current rate of carbon release due to human behaviours is unprecedented \citep{Zeebe2016} and impacts may be felt until at least the year 2500 \citep{Lyon2021, Meinshausen2020}. The Paris Agreement sets a goal of keeping the global average temperature to less than 2$\degree$C above pre-industrial times and aims to limit this increase to 1.5$\degree$C \citep{Paris2015}. Current mitigation strategies are insufficient to limit warming to 2$\degree$C warming \citep{Leach2018, Roelfsema2020} and even meeting these goals is unlikely to prevent sea level rise of at least 1.5 metres by 2300 \citep{Mengel2018} or save coral reefs \citep{Frieler2012}. Clearly, climate action is necessary, but it is not always clear what will shift a population into action and there are many interacting factors \citep{vonBorgstede2013, Brulle2012, gould:2024, cologna:2025, smith:2025, bechthold:2025}. Mathematical modelling provides an avenue to predict and study potential mitigation strategies to slow and reverse climate change \citep{Kemp2022, Beckage2018, Milinski2008, Menard2021, Tavoni2011, Ghidoni2017, Bury2019, Moore2022, savitsky:2025, smith:2025}.

%WHY WE NEED SOCIAL-CLIMATE MODELS--------------------------------------------------------------
The Intergovernmental Panel on Climate Change (IPCC) was created to assess and guide research on climate change through regular assessment reports \citep{IPCC_data}. Many climate models, including those from the IPCC, assume that humans follow a fixed, pre-determined behaviour \citep{Lyon2021, IPCC_data, Meinshausen2020}. That is, researchers make the modelling assumption that humans do not update their behaviours dynamically in response to the changing climate. As a result, and due to the dynamic nature of interactions between the Earth's climate and public opinion, there has been a call from modellers to include a social-climate feedback loop in climate models \citep{Long2014, Beckage2020}. In particular, it is known that human activity is a driving force for climate change \citep{IPCC_data} and many extreme weather events are either directly caused, or worsened, by climate change \citep{Buttke2023, Cai2014, Domeisen2022, Westra2014, Martel2021, Mokria2017}. It is also known that extreme events impact human perception of climate change risk \citep{Bergquist2019, Spence2010} and mitigation behaviour \citep{Demski2016, gould:2024, cologna:2025}. Modelling the climate in this way allows modellers to account for the effects of changes in human behaviour \citep{Beckage2018, Andersson2021, Milinski2008, Menard2021, Tavoni2011, Ghidoni2017, kumar:2025, smith:2025}. In this article, we couple a model of social dynamics with a simple climate model in a feedback loop in order to study the interplay between human behaviour and the climate. Below, we provide a general discussion of the three system components: (1) the social dynamics, (2) the climate dynamics, and (3) the coupling of the two.

%SOCIAL MODELS------------------------------------------------------------------------
\subsection{Social Dynamics}
To include dynamic human behaviour in our model, we first look to the fields of sociology and opinion dynamics. People have a large influence on each other that is often underestimated \citep{Bohns2013, Flynn2008, Melnyk2021, Morris2015}. This influence can take many forms, e.g., propagating social conventions \citep{Melnyk2021, Morris2015}, asking for a favour \citep{Flynn2008}, or even encouraging unethical behaviour \citep{Bohns2013}. Social conventions define the expected behaviours of people participating in society \citep{McDonald2015} and have immense influence on human behaviours such as participation by women in paid labour \citep{Jayachandran2021}, willingness to vaccinate \citep{Graupensperger2021}, decisions on which products to purchase and consume \citep{Melnyk2021}, and level of prejudice \citep{McDonald2015}. 

It has been found that a committed minority can have a profound impact on shifting public opinion and upsetting social conventions \citep{Bolderdijk2021, Moscovici1980}. In particular, the authors suggest that minorities often initiate social change \citep{Bolderdijk2021}, can exert the same amount of influence as a majority population, or may even have a larger proportional effect on the population \citep{Moscovici1980}. Several mathematical models have also studied the ability of a committed minority to overturn social conventions \citep{Galam2007, Xie2011, Centola2018, Wyse2024_social}. These models find that there is a critical size of committed minority within a population, i.e., a threshold below which there is persistence of the social convention, and above which the social convention is overturned.  The latter results in population-level consensus on the opinion held by the committed minority.

%CLIMATE MODELS------------------------------------------------------------------------
\subsection{Climate Systems}
A comparatively simple model of the earth's climate can be formulated using ordinary differential equations (ODEs) \citep{Budyko1969, Geoffroy2013, Luke2010, GlobalCO2Model}. These ODE models can be modified to include stochastic effects, in which case they become stochastic differential equations (SDEs) \citep{Benth2005, Moreles2016, vanderBolt2018}. ODE and SDE models assume that the Earth is is a well-mixed system, i.e., they assume that the climate is the same at every location on the Earth's surface. This simplifying assumption allows us to study the average climate behaviour and gain an understanding of the mechanisms that may be driving more complex models \citep{Wimsatt2007, Epstein2008, Smaldino2017, Haefner2005}. To add complexity and account for spatial effects, some models use partial differential equations \citep{Gordon2000, Hasselmann1976}, or divide Earth into compartments (e.g., lower vs. upper atmosphere) with the assumption that these compartments may behave differently from each other \citep{GlobalCO2Model}. 

Many climate models describe stochasticity using a white noise process that some researchers call a ``mathematical idealization" \citep{vanderBolt2018}. There has since been a shift towards using pink or red noise, i.e., autoregressive noise, to account for temporal and spatial autocorrelation in the climate \citep{DiCecco2018}. In particular, studies have found that temporal autocorrelation, or climate memory, is increasing over time.  This change corresponds to a reddening (i.e., tendency towards higher autocorrelation) of climate stochasticity for various climate metrics \citep{vanderBolt2018, DiCecco2018}. Increased autocorrelation means the climate is more likely to persist in its current state. If this current state involves poor environmental conditions, the climate system may cross a tipping point which may be irreversible \citep{vanderBolt2018}. One study found that the duration of extreme weather events increases as a function of the autocorrelation in the system \citep{vanderBolt2018}. One way to include this pink or red noise in climate models is through the use of SDEs in which the noise is modelled as an Ornstein-Uhlenbeck process \citep{Benth2005, Moreles2016}.

%SOCIAL-CLIMATE MODELS------------------------------------------------------------------------
\subsection{Social-Climate Models}
Modelling the climate using a social-climate feedback loop allows modellers to account for changes in human behaviour in response to extreme events \citep{Beckage2018, gould:2024, cologna:2025}. Other social-climate models study a variety of factors such as, e.g., cooperation to conserve a common resource \citep{Andersson2021, Milinski2008}, unequal distribution of resources impacting the ability to participate in climate mitigation strategies \citep{Menard2021, Tavoni2011, savitsky:2025}, or the impact of delayed damages from emissions \citep{Ghidoni2017}. In general, these models find that factors that delay or prevent climate action, e.g., climate change denial rumours \citep{kumar:2025}, financial inequality \citep{Tavoni2011}, polarization of opinions \citep{Menard2021}, lack of climate policy \citep{vonBorgstede2013, White2019, Andersson2021}, or loss of interest in maintaining climate change mitigation behaviours when there are not immediate climatic improvements \citep{White2019, Christensen2020}, result in higher future temperatures. On the other hand, factors that encourage climate action, e.g., shifting attitudes and social conventions \citep{vonBorgstede2013, White2019, Beckage2018, kumar:2025, bechthold:2025}, co-benefits of action \citep{Bain2015, Marshall2023}, or emphasis on helping future generations \citep{Marshall2023}, result in lower future temperatures and can begin to reverse climate change.

%ORGANIZATION------------------------------------------------------------------------
\subsection{Outline}
The manuscript is outlined as follows. In Section \ref{Sect:Models}, we introduce a social model, a climate model, and then combine them to produce a coupled social-climate model. In Section \ref{Sect:Results}, we describe our climate model results. Then, we present four representative sample simulations of our social-climate model, and the average behaviour observed across all of our social-climate model simulations. We find that a committed minority plays a vital role in shifting public opinion towards climate action and that the time at which the climate inaction social convention is overturned has a large impact on future temperatures. In Section \ref{Sect:Discussion}, we discuss our findings, extensions, and applications of this work. 
%We note that the content from this manuscript was first presented in Chapters 3 and 4 of the first author's MSc thesis \citet{Wyse2024_thesis}. 

\section{Models} \label{Sect:Models}
%-------------------------------------------------------------------------------------
In this manuscript, we combine a social model and a climate model to create a social-climate feedback loop. We propose this toy model as a framework to guide future models. In particular, we choose parameter values in such a way as to provide insight into the types of behaviours we might expect from a more complex model whose parameters are fitted to data. Unless otherwise stated, parameters are chosen so that there is a social model tipping event in about half of the model simulations. We emphasize that the results in this manuscript are qualitative and we focus on the general model behaviour rather than the quantitative aspects of the results. 

\subsection{Social Model} \label{Sect:Models_Social}
%-------------------------------------------------------------------------------------

The social model we use is based on the social agent-based model (ABM) presented in \citet{Centola2018}. We are interested in investigating the influence that a small proportion $\CM$ of committed individuals, i.e., a committed minority, can have on a population. The remaining proportion $1-\mathcal{C_M}$ of the population is uncommitted to any particular opinion, i.e., individuals in this part of the population are willing to change their opinion. Opinions $A$ and $B$ spread throughout the uncommitted proportion of the population and evolve when individuals participate in pairwise speaking-listening interactions. The individuals in this social model have memory banks in which they hold memories of their previous $M$ listening interactions. In each interaction, the speaker speaks their opinion, i.e., the most frequent memory in their memory bank, and, if the listener is uncommitted, the listener adds this memory to their memory bank in a first-in-first-out process. Without loss of generality, we set $A$ as the opinion of individuals in the committed minority and $B$ as the opinion initially held by everyone else.

Since changes in the climate are slow compared to changes in opinions \citep{Ricke2014}, we will find it useful to assume that the social dynamics are always at steady-state.  In \citet{Wyse2024_social}, we discuss the mean-field approximation of this model and obtain the opinion response function (ORF) which can be used to determine the steady state behaviour of the social model. Here we summarize the main points. The ORF takes, as its input, a constant rate at which individuals in the population are subjected to listening to each opinion, and returns, as its output, the steady-state speaking rate of each opinion. A steady state then corresponds to a fixed point of the ORF, since for each opinion, the speaking and listening rate are the same; moreover, for any initial value, repeated application of the ORF leads to the same steady state as the one obtained by simulating the social model. Since, in this model, there are two opinions, whose per-capita speaking rate sums to $1$, it suffices to use a single parameter $r$ that represents the rate of listening to, or speaking, opinion $A$. By iterating the ORF to a fixed point, we can implement the steady-state assumption for the social model as $r$ changes, due to, e.g., an extreme climatic event, in a way that is more computationally efficient than simulating the slower and more complex ABM. %\sout{corresponding $2^M$-dimensional set of ordinary differential equations. The ORF calculation is much faster and simpler than the ABM simulation.} 

The ORF %considers speaking and listening as two decoupled processes \citep{Diekmann2003} and 
is given by 
\begin{equation}
    \Psi_\mathcal{C_M}(r) = \mathcal{C_M} + (1-\mathcal{C_M})\Phi(r).
\end{equation}
Here, $\Psi_\mathcal{C_M}(r)$ is the steady-state speaking rate of opinion $A$ in a population that hears opinion $A$ at constant rate $r$, in which the committed minority speaks opinion $A$ in every speaking interaction. The uncommitted population speaks $A$ at rate $\Phi(r)$ which depends on the configuration of the memory banks of individuals in the uncommitted population. Hearing opinion $A$ at rate $r$ and opinion $B$ at rate $1-r$, at equilibrium, in each position of its memory bank, an uncommitted individual holds $A$ with probability $r$, so % , an uncommitted individual To determine $\Phi(r)$ for the uncommitted population, we start by considering an uncommitted individual. We define $X(r)$ as the memory bank for an individual in the uncommitted population as a function of $r$, the hearing rate of opinion $A$. We note that if 
the number of $A$ memories in the individual's memory bank is binomial with parameters $M,r$. Denoting this random variable by $N_r$, %an individual's memory bank is greater than half of their available memory slots, i.e., 
if $N_r>M/2$, then the individual will hold and speak opinion $A$, %. If an individual holds exactly half $A$ memories and exactly half $B$ memories, i.e., 
and if $N_r= M/2$, then the individual is undecided and speaks opinion $A$ in half of their speaking interactions. %To obtain the speaking rate of opinion $A$ for the uncommitted population, we compute 
Since each individual speaks at rate $1$, the rate at which an uncommitted individual speaks opinion $A$ is equal to the probability that it speaks $A$ on a given speaking event, therefore %individuals in the uncommitted population hold a memory bank with more than half $A$ memories or exactly half $A$ memories. Altogether, the speaking rate of opinion $A$ in the uncommitted population is 
\begin{equation}
    \Phi(r) = P(N_r>M/2)+\frac{1}{2}P(N_r= M/2).
\end{equation}
As mentioned above, the fixed points of $\Psi_\mathcal{C_M}(r)$ correspond to the steady states of the mean-field approximation of the social model. Furthermore, there exists $\mathcal{C_M}^*$ such that there is a saddle node bifurcation when $M\geq3$. For $\mathcal{C_M}<\mathcal{C_M}^*$, there is coexistence of both opinions and for $\mathcal{C_M}>\mathcal{C_M}^*$, there is consensus on opinion $A$. We refer readers to \citet{Wyse2024_social} for further details on our social model.

\subsection{Climate Model} \label{Sect:Models_Climate}
%-------------------------------------------------------------------------------------
A commonly used non-spatial energy-balance model for the Earth describes energy entering the system via absorbed solar radiation (ASR) and energy leaving the system via outgoing longwave radiation (OLR) \citep{Tung2007}. Multiplying these energy terms by the specific heat capacity of water, $C$, produces an equation that models the temperature $T$ of the Earth's surface over time, 
\begin{equation}
\label{Eq:Clim:EBM}
	C\frac{dT}{dt} = \frac{S(1-\alpha)}{4} - \tau \sigma T^4,
\end{equation}
where $S(1-\alpha)/4$ is the ASR and $- \tau \sigma T^4$ is the OLR \citep{Tung2007}. Here, $\tau$ is the transmissivity of the atmosphere and $\sigma$ is the Stefan-Boltzmann constant. The remaining parameters are described in Table \ref{Tab:Clim:Params}. 

\begin{sidewaystable}[tbph]
%\begin{landscape}
%\begin{table}[tbph]
\caption{Variables and parameters used in the climate model. Note that while $C$ is given in unit of seconds (J=W*seconds), we make a conversion to years to match the IPCC $CO_2$ emission projections.}
\begin{tabular}{lllll}
	Parameter & Definition & Value & Units & Source \\
	\hline
	$T$ & temperature & variable & $\degree$C & \\
	$T_{PI}$ & pre-industrial temperature & 13.7 & $\degree$C & \citep{Lenssen2019} \\
	$T_{2020}$ & 2020 temperature & 14.9 & $\degree$C & \citep{IPCC_data} \\
	$t$ & time & variable & years & \\ 
	$C$ & specific heat capacity of ocean water  & $3.985\times 10^6$ & J/m$^2$$\degree$C & \cite{seaWater_heatCapacity} \\
	  & \qquad to a depth of 50 m  & & & \\
	$S$ & solar radiation & 1372 & W/m$^2$ & \cite{Tung2007} \\
	$\alpha$ & albedo & 0.33 & unitless & \cite{Tung2007} \\
	$\eta$ & climate feedback parameter & 202 & W/m$^2$ & \cite{Tung2007} \\
	$\beta$ & climate feedback parameter & 1.9 & W/m$^2$$\degree$C & \cite{Tung2007} \\
	$a$ & CO$_2$ forcing coefficient & 5.35 & W/m$^2$ & \cite{Myhre1998} \\
	CO$_2$ & projected CO$_2$ atmospheric concentration & variable & Gt/year & \cite{IPCC_data} \\
	CO$_{2,PI}$ & pre-industrial CO$_2$ atmospheric & 280 & ppm & \cite{IPCC_CO2conc} \\
	  & \qquad concentration  & & & \\
	CO$_{2,2020}$ & 2020 CO$_2$ atmospheric concentration & 410 & ppm & \cite{IPCC_CO2conc} \\
	\hline
\end{tabular}
\label{Tab:Clim:Params}
%\end{table}
%\end{landscape}
\end{sidewaystable}

This model~\eqref{Eq:Clim:EBM} does not account for the impact of greenhouse gas (GHG) emissions on parameters such as the transmissivity of the atmosphere. In particular, as GHGs increase, the atmospheric transmissivity decreases \citep{Rose2022}. To include the impact of GHG emissions in our model, the OLR term in~\eqref{Eq:Clim:EBM} can be linearized and formulated in terms of emissions rather than temperature. We follow \citet{Tung2007} and obtain
\begin{equation}
\label{Eq:Clim:EBM_GHG}
	C\frac{dT}{dt} = \frac{S(1-\alpha)}{4} - (\eta -\beta T) + a\log{\left(\frac{\text{CO}_2(t)}{\text{CO}_{2,PI}}\right)}.
\end{equation}
The ASR term remains the same as in Equation \ref{Eq:Clim:EBM}. The term $\eta-\beta T$ represents energy lost in outgoing radiation where both $\eta$ and $\beta$ are climate feedback parameters.  Different sources provide different values for $\eta$ and $\beta$, and these values change across the globe \citep{Graves1993, Tung2007, Meier2020}. We use the empirically-based values from \citet{Graves1993}. 

The last term describes the temperature forcing from current CO$_2(t)$ emissions as compared to pre-industrial levels, CO$_{2,PI}$, where CO$_2(t)$ is measured in parts per million (ppm). We use CO$_2$ emissions projections (Table \ref{Tab:Clim:CO2EmisConc}, \citet{IPCC_data}) from the IPCC. The IPCC studies five ``high priority" climate scenarios under the shared socio-economic pathways (SSPs) framework \citep{Meinshausen2020}. The five scenarios they consider are SSP1-1.9, SSP1-2.6, SSP2-4.5, SSP3-7.0, and SSP5-8.5. The first number represents the SSP family where SSP1 is called sustainable development, SSP2 is middle-of-the-road development, SSP3 is regional rivalry, SSP4 is inequality, and SSP5 is fossil-fuel development. The rest of the numbers describe the radiative forcing level by 2100 given in W/m$^2$. For example, SSP1-1.9 is in the sustainable development family and predicts 1.9W/m$^2$ radiative forcing in 2100. The emissions projections from these five scenarios are given in gigatonnes per year for each decade. In order to obtain annual emissions for our model, we first linearize between the given data values within each climate scenario \citep{IPCC_data} and then convert the data to units of ppm. Lastly, we compute the current CO$_2(t)$ concentration in the climate model under each scenario by summing the 2020 CO$_2$ concentration with the emissions up to the current time step. 

\begin{table}[tbph]
\centering
\caption{Projected CO$_2$ emissions, in gigatonnes per year, for each IPCC scenario, and given in 10 year increments. Data from \cite{IPCC_data}.}
\begin{tabular}{llllll}
	Year & SSP1-1.9 & SSP1-2.6 & SSP2-4.5 & SSP3-7.0 & SSP5-8.5\\
	\hline
	2020 & 39.69 & 39.80 & 40.65 & 44.81 & 43.71 \\
	2030 & 22.85 & 34.73 & 43.48 & 52.85 & 55.30 \\
	2040 & 10.48 & 26.51 & 44.25 & 58.50 & 68.78 \\
    2050 &  2.05 & 17.96 & 43.46 & 62.90 & 83.30 \\
    2060 & -1.53 & 10.53 & 40.20 & 66.57 & 100.34 \\
    2070 & -4.48 & 4.48	& 35.24 & 70.04 & 116.81 \\
    2080 & -7.31 & -3.29 & 26.84 & 73.41 & 129.65 \\
    2090 & -10.57 & -8.39 & 16.32 & 77.80 & 130.58 \\
    2100 & -13.89 & -8.62 & 9.68 & 82.73 & 126.29 \\
	\hline
\end{tabular}
\label{Tab:Clim:CO2EmisConc}
\end{table}

To include climatic variation in the global average temperature in our model, we modify~\eqref{Eq:Clim:EBM_GHG} by adding a stochastic term as follows:
\begin{subequations}
\label{Eq:Clim:EBM_SDE}
\begin{align}
	dT &= \left(\frac{S(1-\alpha)}{4} - (\eta -\beta T) +a\log{\left(\frac{\text{CO}_2(t)}{\text{CO}_{2,PI}}\right)}\right)\frac{dt}{C} + dW, \label{Eq:Clim:EBM_SDE_1} \\
    dW &= -\theta W dt + \sigma dB. \label{Eq:Clim:EBM_SDE_2}
\end{align}
\end{subequations}
Equation \eqref{Eq:Clim:EBM_SDE_2} defines an Ornstein-Uhlenbeck process $W$. Here, $\theta =1/5$ corresponds to global average temperature having a mean-reverting time of 5 years and $\sigma = 1$ \citep{Hansi2023}. Since $0<\theta<\infty$, our SDE formulation is an energy balance model with additive pink noise \citep{Mustin2013}. In this model, $dB=\sqrt{dt}Z$ is a Gaussian random variable where $Z\sim N(0,1)$.

\subsection{Social-Climate Model} \label{Sect:Models_SCM}
%-------------------------------------------------------------------------------------

\begin{figure}[tbph]
    \centering
    \includegraphics[width=\textwidth]{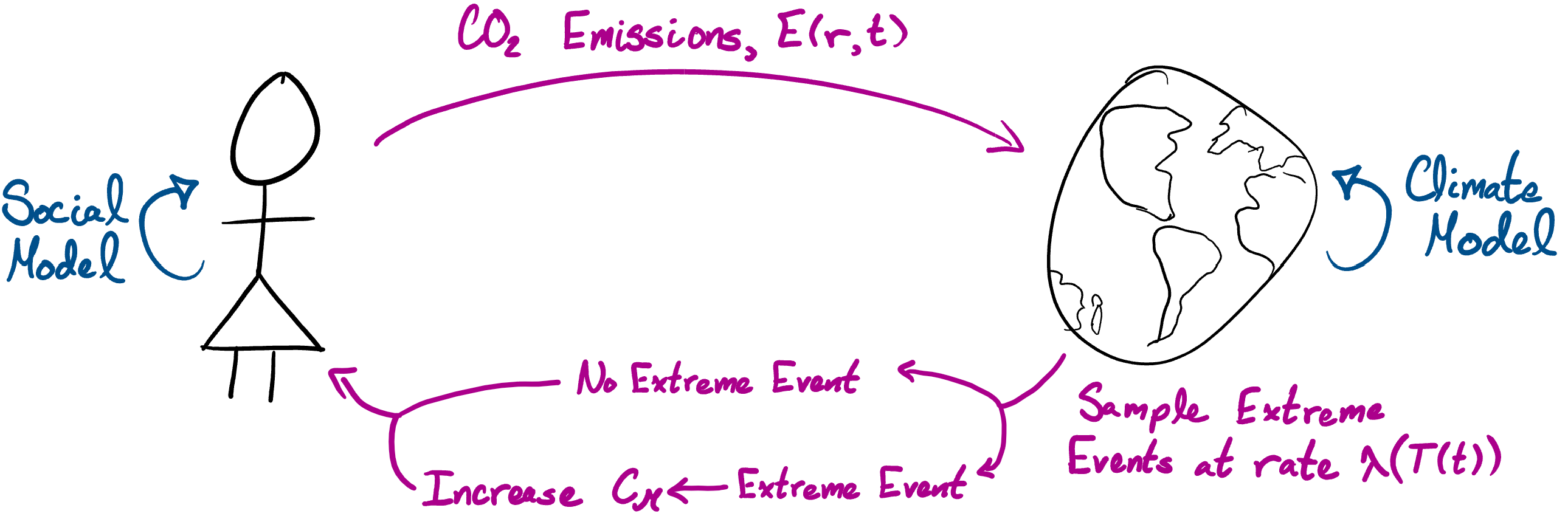}
    \caption{An outline of our social-climate model feedback loop.}
    \label{fig:outline}
\end{figure}

The novel contribution of this manuscript is to connect our simple social model, which includes committed minority dynamics, with a simple stochastic climate model in a social-climate feedback loop (Figure \ref{fig:outline}). In this section we explain how the two sub-models are connected.

We start by building the social-to-climate model connection (i.e., the upper arrow in Figure \ref{fig:outline}). At each time step, we use the equilibrium speaking rate of the climate action opinion in the social model, $r(t)$, to determine the level of current CO$_2$ emissions, 
$E(t)$, 
which is a weighted sum of the emissions expected under the worst, $E_{\text{worst}}$, and best, $E_{\text{best}}$, case SSP scenarios, given in gigatonnes per year. At every time step, SSP1-1.9 represents the best-case emissions and SSP5-8.5 represents the worst-case emissions. We obtain: 
\begin{equation}
\label{eq:Emis}
    E(t) = r(t)E_{\text{best}}(t) + (1-r(t))E_{\text{worst}}(t).
\end{equation}
We follow the same steps as in Section \ref{Sect:Models_Climate} to convert these CO$_2$ emissions into a current CO$_2(t)$ concentration in ppm. We feed this concentration into Equation \ref{Eq:Clim:EBM_SDE} and run one time step (one day) of the climate model.

Next, we build the climate-to-social model connection (i.e., the lower arrows in Figure \ref{fig:outline}). We set $\lambda_0=4$ as the base rate of extreme weather events in the absence of climate change \citep{CarbonBrief2022}. In general, models predict that extreme events will increase in frequency and intensity as mean global temperature increases \citep{IPCC_data, Cai2014, Martel2021, Buttke2023, Mokria2017}. The Clausius-Clapeyron relationship predicts that extreme rainfall will increase by 7\% for each 1$\degree$C increase in global average temperature \citep{Martel2021}. An empirical study suggests that extreme rainfall will increase by as much as 14\% as global warming progresses \citep{Westra2014}.  We choose the approximate midpoint and use a 10\% increase in extreme precipitation events, per degree Celsius, as a proxy for the increase in frequency of all extreme weather events (e.g., extreme heat, extreme drought, hurricanes). Hence, we take the rate of extreme events, $\lambda(T(t))$, to be
\begin{equation}
\label{eq:SCM:rateEE}
    \lambda(T(t)) = \lambda_0 1.1^{T(t)-T_{PI}},
\end{equation}
where $T_{PI}$ and $T(t)$ are the global average surface temperatures from pre-industrial times and at time $t$, respectively. 

To determine the number of extreme events at each time step, we sample from a Poisson distribution with rate $\lambda(T(t))$. If our sampling procedure returns zero extreme events, we return to the social model and repeat the feedback loop. If our sampling procedure returns an extreme event, then we increase the speaking rate of the climate action opinion in the social model by increasing $\mathcal{C_M}(t)$. We note that increasing $r(t)$ directly is ineffective as any increase would be undone once the simulation returns to the social model (via the lower arrow in Figure~\ref{fig:outline}) since the first step in the social model is computing the equilibrium $r(t)$ based on the committed minority size (see \citet{Wyse2024_social}). Instead, by increasing $\mathcal{C_M}(t)$ during the climate to social connection step, we indirectly increase $r(t)$ in the social step. We let this increase in $\mathcal{C_M}$ decay over time to account for extreme events decreasing in impact over time \citep{Ray2017, Walshe2020}. The size of the committed minority is therefore given by the decaying cumulative sum
\begin{equation}
\label{eq:shiftCM}
    \mathcal{C_M}(t) = \mathcal{C_M}(0) + \sum_{s\in EE} \mu \exp{(-\delta(t-s))},
\end{equation}
where $\mu=0.014$ is the size of the temporary increase to the committed minority, $\delta=0.002$ is the decay rate of this increase, and $EE$ is the set of time indices for each extreme event up until time $t$. We assume that the impact of extreme events that occur before our simulations start in 2020 are accounted for in the initial condition $\mathcal{C_M}(0)=0.1$. Note that $\mathcal{C_M}(0)$ is also the baseline committed minority size to which $\mathcal{C_M}(t)$ decays as later extreme events are forgotten. We feed this shifted $\mathcal{C_M}(t)$ back into the social model and repeat the procedure until year 2100.

\section{Results} \label{Sect:Results}
%-------------------------------------------------------------------------------------
Results from our social model and opinion response functions were discussed in earlier work \citep{Wyse2024_social}, and so we present just a brief summary of the key results here (Section~\ref{Sect:Results_Social}). We then discuss our findings from our stochastic energy balance climate model (Section~\ref{Sect:Results_Climate}) and our social-climate model (Section~\ref{Sect:Results_SCM}).

\subsection{Social Model Results} \label{Sect:Results_Social}
%-------------------------------------------------------------------------------------

\begin{figure}[tbph]
    \centering
    \includegraphics[width=0.5\textwidth]{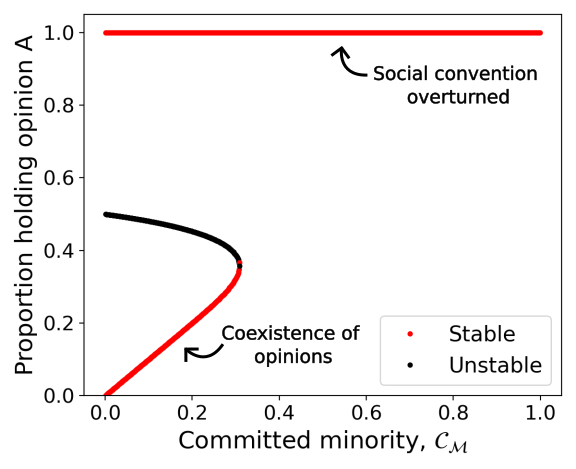}
    \caption{A sample bifurcation diagram for our social model with $M=25$. 
%    \sw{If we want the "coexistence of opinions" and "social convention overturned" labels removed, I can do that. I didn't want to reuse the exact same figures from our social model paper, so I found one I had used on my SCM poster.}
    }
    \label{fig:Social_SampleBifur}
\end{figure}

In our social model, we find two types of model behaviour for $M<3$ and $M\geq3$. Since we use $M=25$ in our social-climate model, we will focus on the $M\geq3$ results here. In this case, there is a non-trivial saddle-node bifurcation at $\mathcal{C_M}^*$ (see sample a bifurcation diagram in Figure \ref{fig:Social_SampleBifur}). For $\mathcal{C_M}<\mathcal{C_M}^*$, the opinion $B$ social convention persists and only a small proportion of the population holds opinion $A$. However, when the committed minority population increases to $\mathcal{C_M}>\mathcal{C_M}^*$, the social convention is overturned and the only remaining steady state is consensus on opinion $A$. We note that this bifurcation is irreversible since even if the committed minority decreases, there is no longer anyone to sway the population to revert back to opinion $B$. 

%the relationship between M and C_M^* is non-decreasing (i.e., increases in M lead to the same or increases C_M^* - could be included but does this result matter to the SCM???

\subsection{Climate Model Results} \label{Sect:Results_Climate}
%-------------------------------------------------------------------------------------

The initial conditions we use for our climate model (Equations \ref{Eq:Clim:EBM_SDE}) are obtained from temperature and CO$_2$ concentration data from 2020 \citep{IPCC_data, IPCC_CO2conc}. We use projections of future CO$_2$ emissions (Table \ref{Tab:Clim:CO2EmisConc}) from the IPCC to calculate projections for CO$_2$ concentration levels for the five IPCC climate scenarios \citep{IPCC_data}. Since these projections were only run until 2100, our simulations are also restricted to end in year 2100. At each time step, we sample $dB\sim N(0,1)$ and solve the system using the Euler-Maruyama scheme. 

\begin{figure}[tbph]
    \centering
    \includegraphics[width=0.6\textwidth]{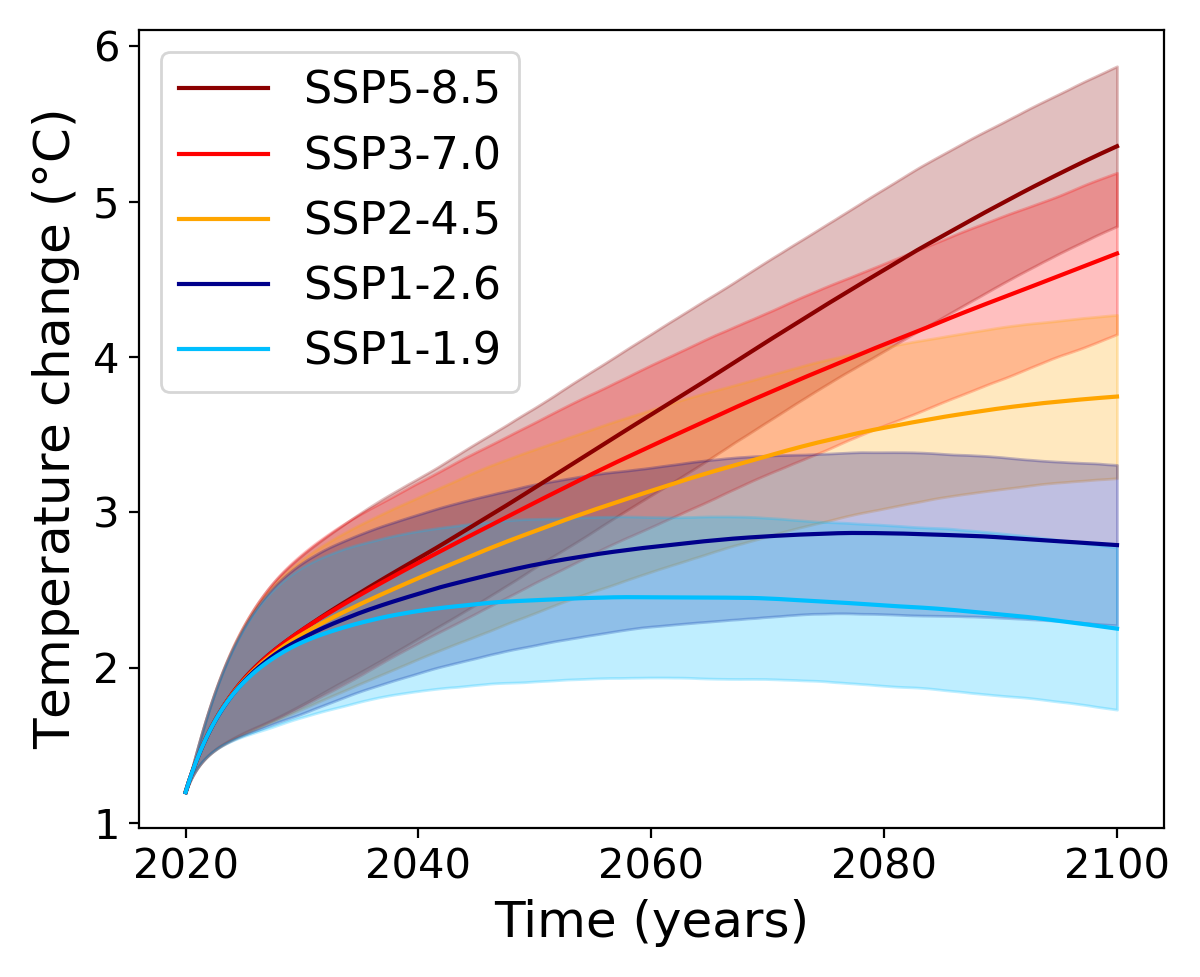}
    \caption{Projections of global surface temperature change from the SDE model for the five IPCC climate scenarios. The solid lines show the average behaviour across 10,000 simulations and the shaded regions surrounding each solid line are the 95\% confidence intervals.}
    \label{fig:Climate_SDE}
\end{figure}

We run 10,000 simulations to obtain the average behaviour and 95\% confidence intervals for each of the five IPCC climate scenarios 
(Figure \ref{fig:Climate_SDE}). Since the average behaviour of the SDE model, for each climate scenario, is the same as that of the corresponding ODE model, we only include the SDE results here. The first ten years of temperature change projections are essentially the same across the five climate scenarios. As time passes, differences in emissions levels have an effect. Scenarios SSP2-4.5, SSP3-7.0, and SSP5-8.5 show increasing temperatures until 2100 with the worst case scenario (SSP5-8.5) reaching an average maximum temperature change of 5.36$\degree$C by 2100. In the best case scenario (SSP1-1.9), the average maximum temperature change is 2.46$\degree$C and occurs in 2059. After that date, the global surface temperature starts decreasing until it reaches an average of 2.25$\degree$C by 2100. Scenario SSP1-2.6 has behaviour that is similar to SSP1-1.9, though the average maximum temperature is greater (2.86$\degree$C), occurs later (year 2079), and the average final temperature change is also greater (2.79$\degree$C). The 95\% confidence intervals initially have zero width since the initial conditions for every simulation are the same. The width increases to about 0.5$\degree$C, after which it remains approximately constant, corresponding to a standard deviation of about 0.25$\degree$C. 

To investigate the accuracy of our model, we compare it to results from the IPCC. We note that our model only uses two equations along with CO$_2$ projections from the IPCC. Our model is significantly simpler than other climate models, yet it produces temperature predictions that compare well with those of the IPCC models (see Figure SPM.8(a) in \citep{IPCC_data}). In particular, the qualitative shapes of our climate model predictions and the IPCC model predictions for years 2030 onward are very similar and we obtain similar temperature predictions for 2100, with our model predicting temperatures approximately 0.5$\degree$C higher for most scenarios. The main difference in model behaviour is the first decade of projections. In this region, our model shows a sharper increase in global surface temperature than the IPCC model. For our purposes, the match between the two models is acceptable.

%It is possible that adjusting parameters, such as the climate feedback parameters $\eta$ and $\beta$, would shift our model behaviour to better align with the IPCC results. That said, simulations using other values of $\eta$ and $\beta$ did not appear to produce a better fit.

\subsection{Social-Climate Model Results} \label{Sect:Results_SCM}
%-------------------------------------------------------------------------------------

To study the conditions under which a committed minority can shift public opinion towards climate action, we choose opinion $A$ as the climate action opinion. We call opinion $B$ the climate inaction or business as usual opinion. We set memory bank length $M=25$, meaning that the social model has a tipping point at committed minority proportion $\mathcal{C_M}^*\approx 0.3086$ \citep{Wyse2024_social}. We choose an initial condition $\mathcal{C_M}=0.1$ for our social model. For the climate model, we use the same initial conditions as for the climate model alone. 

We run 10,000 simulations to investigate the dynamics of our social-climate model. From these, we first focus on four representative sample simulations, then examine the distribution of the final temperature conditioned on whether or not the social model has a tipping event. The four representative simulations are: the simulation yielding (1) the minimum final temperature, (2) a social model tipping event occurring late in the simulation, (3) the maximum final temperature, and (4) the largest maximum temperature. We describe each of these sample simulations in more detail in the following paragraphs. We do not consider the case of the lowest minimum temperature since the minimum temperature in all of our simulations is the initial temperature. Thus, all of our simulations achieve the smallest minimum temperature.

\begin{figure}[tbph]
    \centering
    \includegraphics[width=0.48\textwidth]{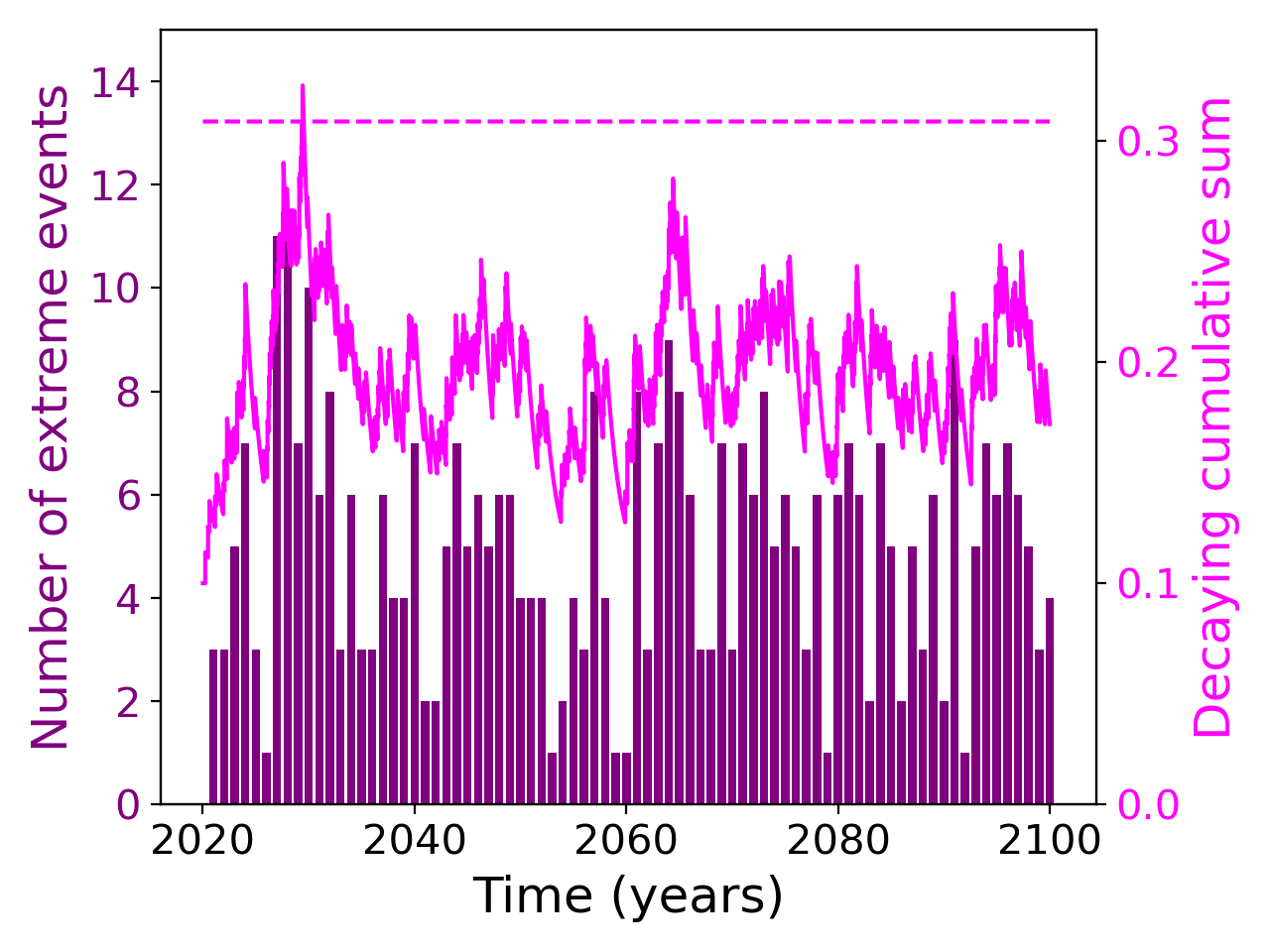}
    \includegraphics[width=0.48\textwidth]{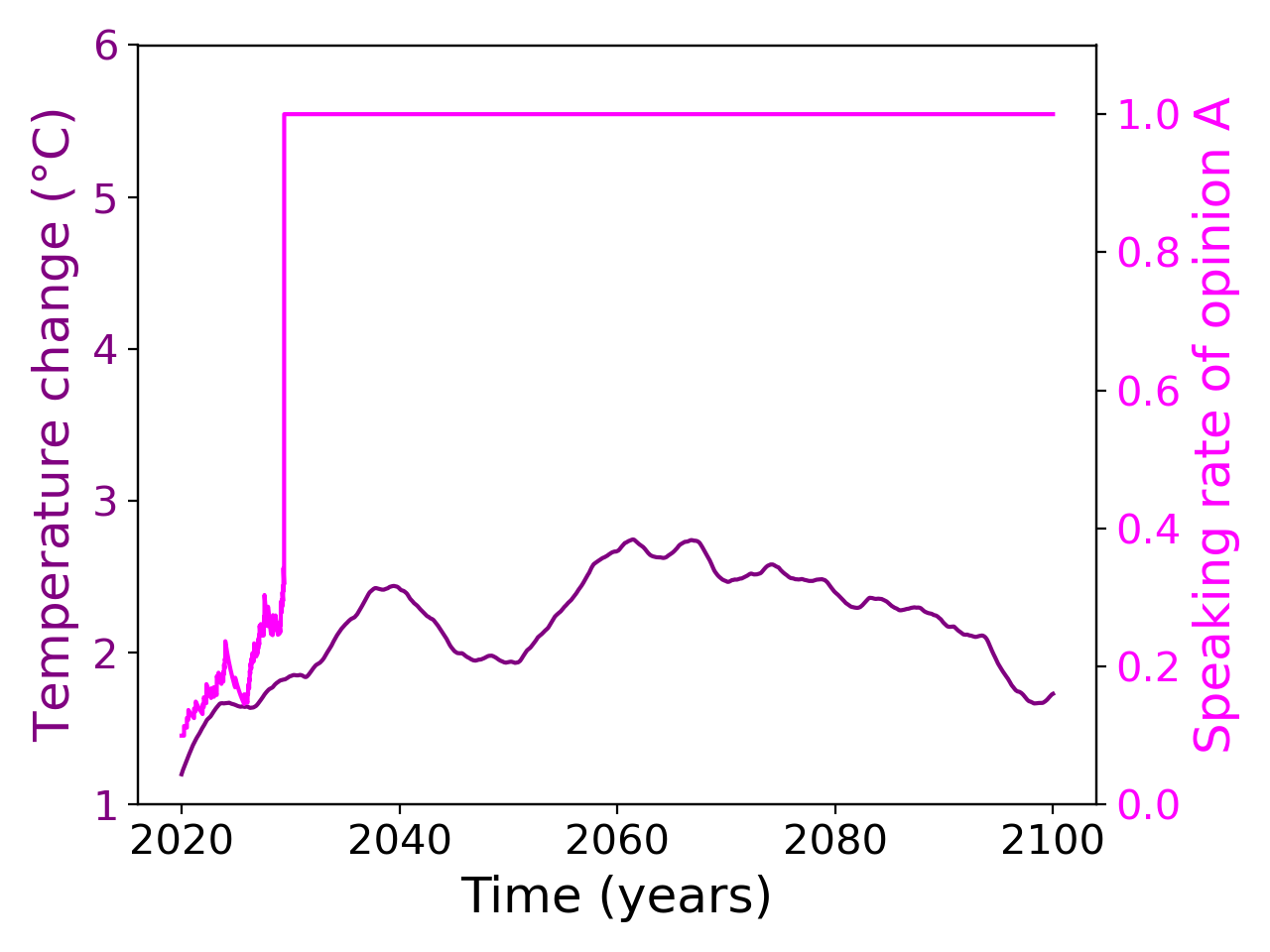}
    \caption{Annual counts of extreme events and decaying cumulative sum (left), speaking rate of $A$, and global surface temperature change from pre-industrial times (right) in the minimum final temperature scenario. The dashed line (left) represents the social model tipping point, $\mathcal{C_M}^*$.}   
    \label{fig:SCM_minTemp}
\end{figure}

The simulation with the minimum final temperature (Figure \ref{fig:SCM_minTemp}) represents the best case scenario from our simulations. We find that a sufficiently high frequency of extreme events over a sufficiently short time interval causes a large enough (temporary) shift of the uncommitted population into the committed minority to cause a tipping event in the social model (i.e., the decaying cumulative sum \eqref{eq:shiftCM} crosses the threshold $\mathcal{C_M^*}$). In this simulation, the high number of extreme events and the tipping event both occur at about 2030. Following this tipping event, there is an immediate shift towards climate action and $r(t)=1$. This shift in opinion means Equation~\eqref{eq:Emis} simplifies to $E(t)=E_{\text{best}}(t)$ and the emissions for the rest of the simulation are exactly the emissions from SSP1-1.9. We note that since the tipping point in our social model is irreversible, there is no shift back towards any individuals speaking $B$ or higher emissions. As a result, extreme events after the tipping event no longer have an effect since everyone in the committed minority and uncommitted population is already speaking opinion $A$.

\begin{figure}[tbph]
    \centering
    \includegraphics[width=0.48\textwidth]{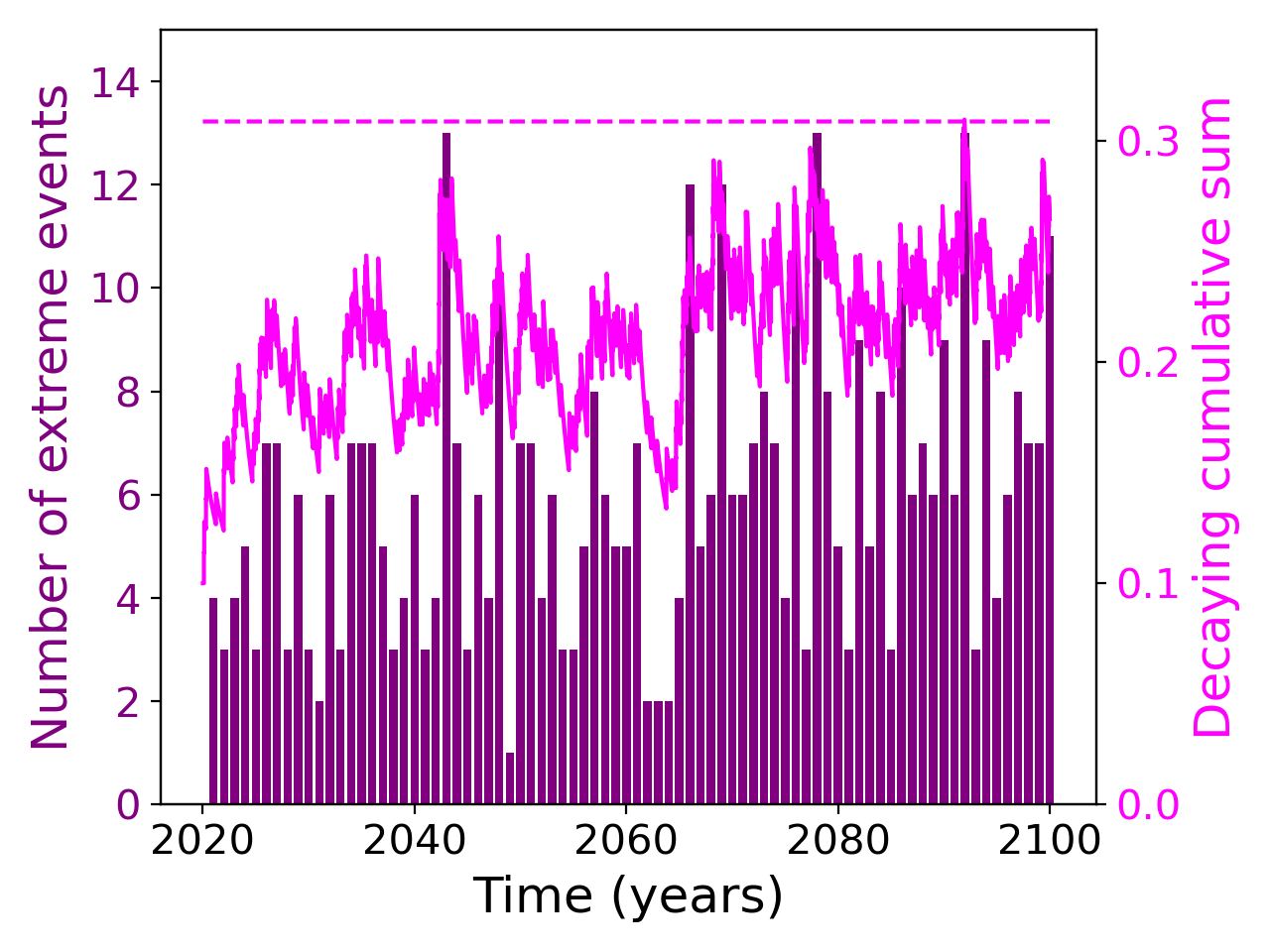}
    \includegraphics[width=0.48\textwidth]{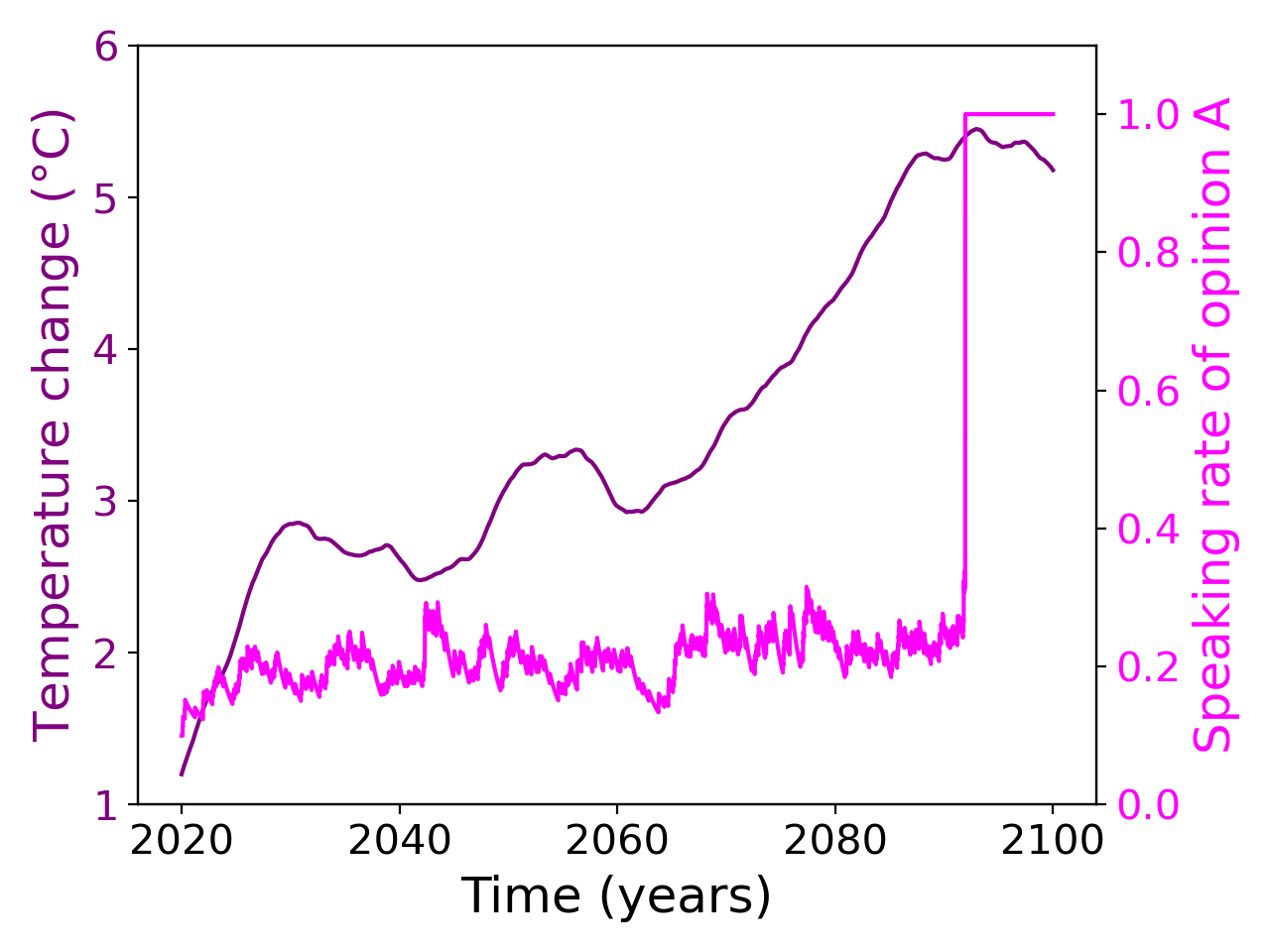}
    \caption{Annual counts of extreme events and decaying cumulative sum (left), speaking rate of $A$, and global surface temperature change from pre-industrial times (right) in the late social model tipping event scenario. The dashed line (left) represents the social model tipping point, $\mathcal{C_M}^*$.}   
    \label{fig:SCM_lateTip}
\end{figure}

When the social model tipping event occurs late in the simulation (roughly 2090 in Figure \ref{fig:SCM_lateTip}), the temperature increase since pre-industrial times is substantial at the tipping time. The remaining 10 years between the tipping event and the year 2100 represents a very short period of decreasing emissions in comparison to the period of increasing emissions pre-tip. Post-tip, there is a substantial drop in emissions from approximately 100 gigatonnes CO$_2$ per year to -10.57 gigatonnes per year within one time step (see SSP1-1.9 and SSP5-8.5 in the second last row of Table \ref{Tab:Clim:CO2EmisConc} and combine with Equation \ref{eq:Emis}). This sudden and dramatic drop is a result of the form of Equation \ref{eq:Emis} and the bifurcation structure of our social model. That is, within one time step, there is a mass shift from $r(t)\approx0.2$ to $r(t+1)=1$ (right plot in Figure \ref{fig:SCM_lateTip}) which results in a large change in emissions. While this change in emissions is immediate, the change in atmospheric concentration of CO$_2$ is not as immediate since the concentration of CO$_2$ decays at maximum rate of -13.89 gigatonnes CO$_2$ per year (see SSP1-1.9 in the last row of Table \ref{Tab:Clim:CO2EmisConc}). 

\begin{figure}[tbph]
    \centering
    \includegraphics[width=0.48\textwidth]{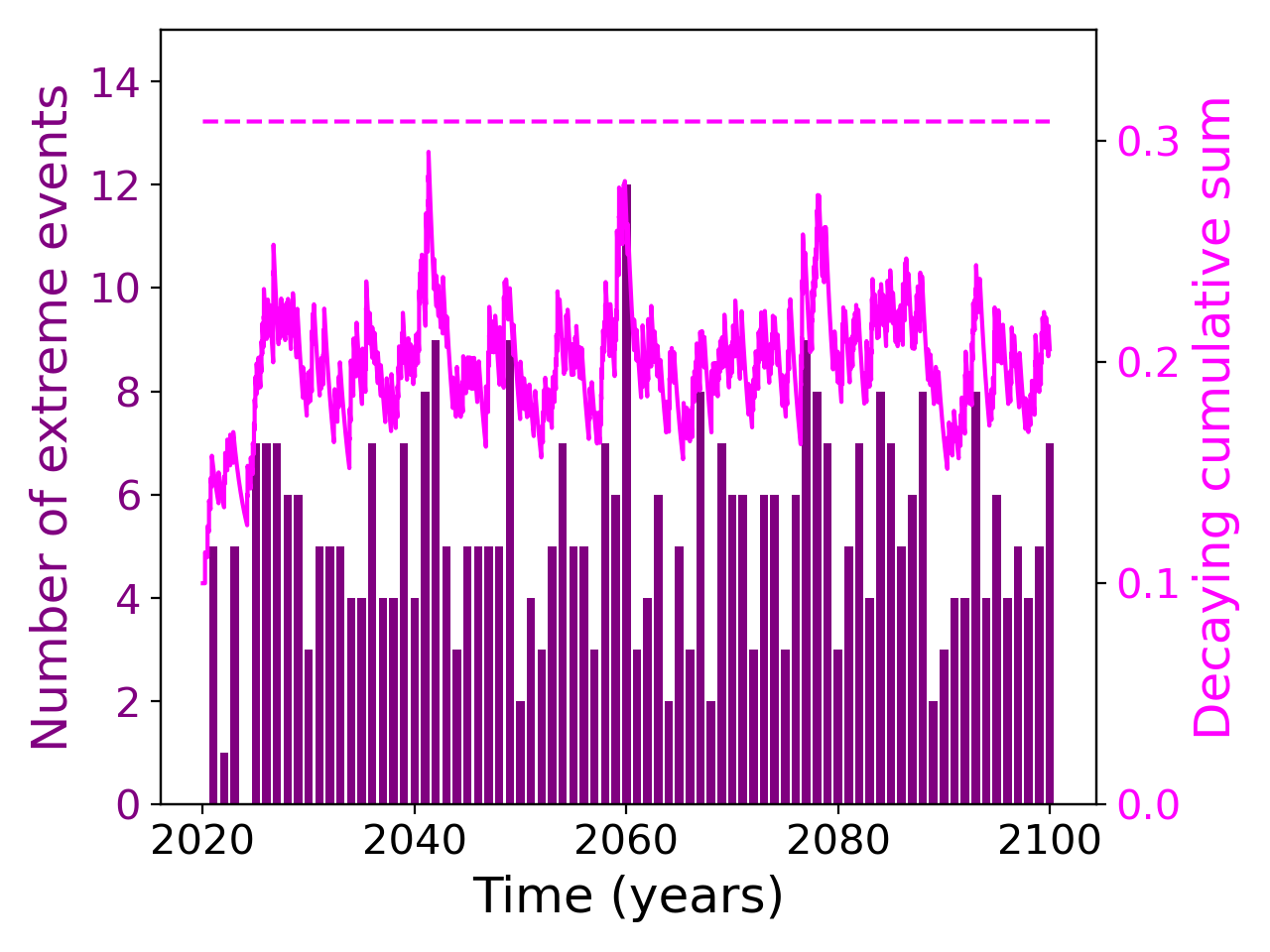}
    \includegraphics[width=0.48\textwidth]{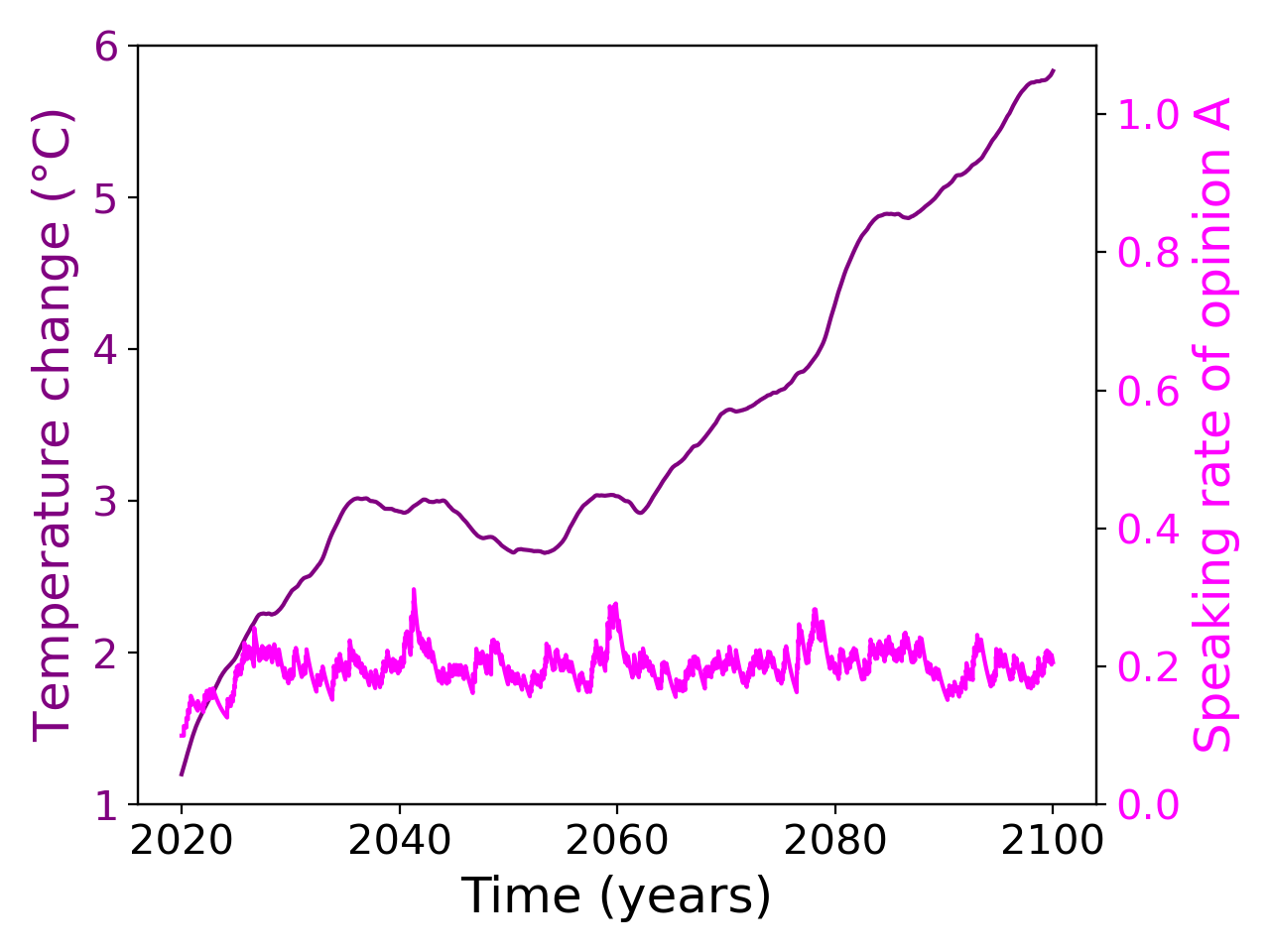}
    \caption{Annual counts of extreme events and decaying cumulative sum (left), speaking rate of $A$, and global surface temperature change from pre-industrial times (right) in the maximum final temperature scenario. The dashed line (left) represents the social model tipping point, $\mathcal{C_M}^*$.}   
    \label{fig:SCM_maxTempEnd}
\end{figure}

\begin{figure}[tbph]
    \centering
    \includegraphics[width=0.48\textwidth]{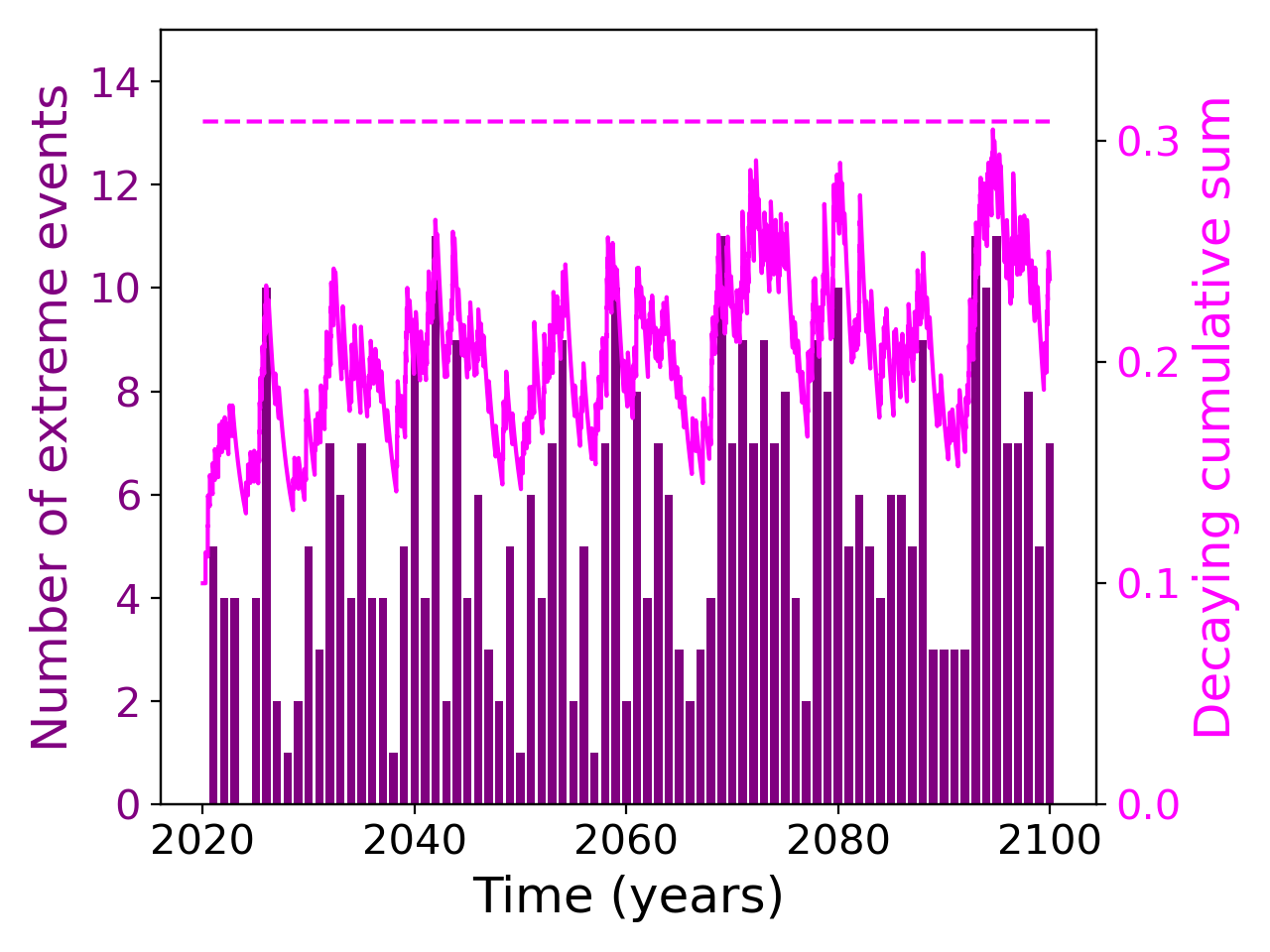}
    \includegraphics[width=0.48\textwidth]{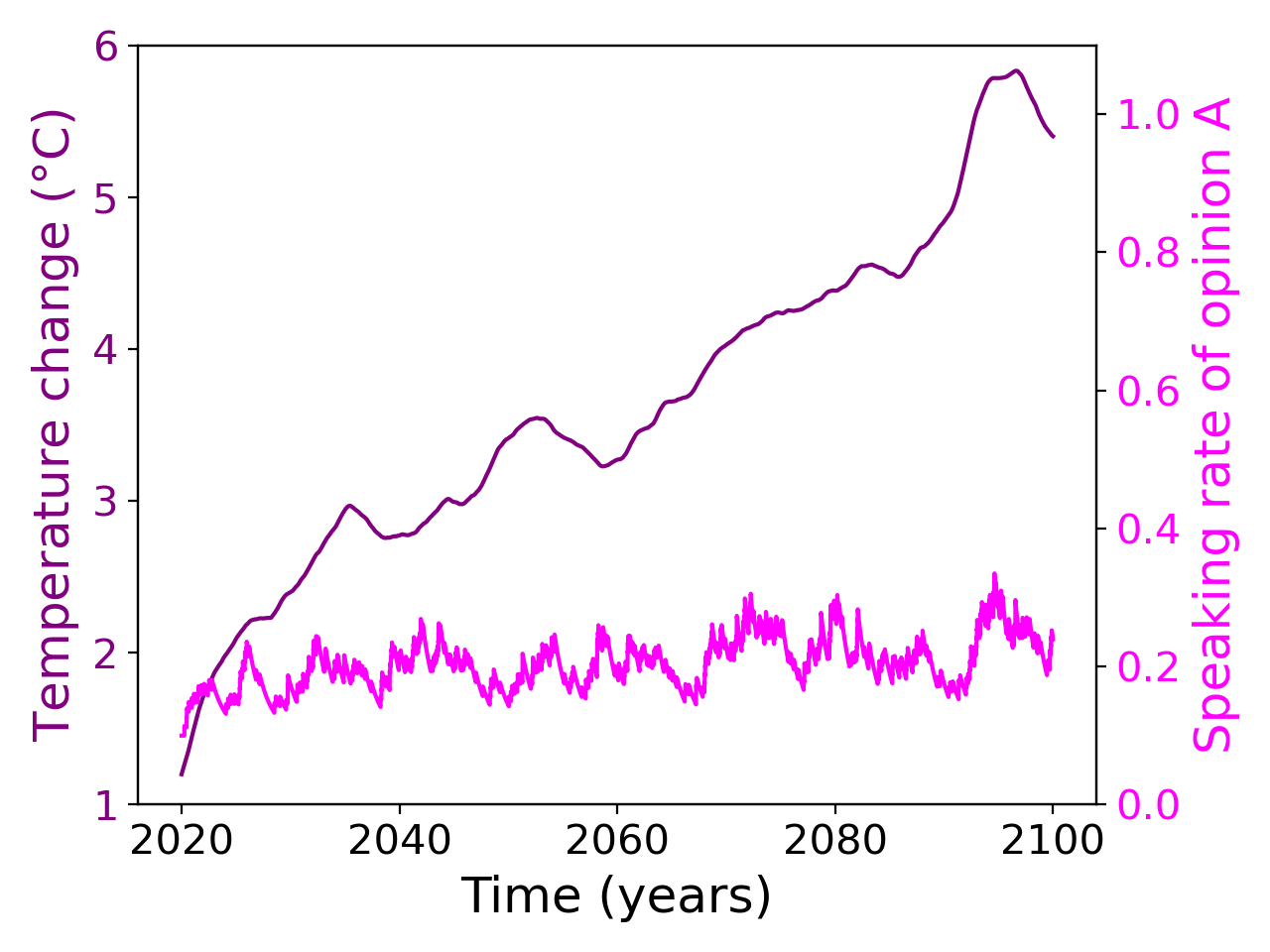}
    \caption{Annual counts of extreme events and decaying cumulative sum (left), speaking rate of $A$, and global surface temperature change from pre-industrial times (right) in the largest maximum temperature scenario. The dashed line (left) represents the social model tipping point, $\mathcal{C_M}^*$.}   
    \label{fig:SCM_maxTempThroughout}
\end{figure}

In the maximum final temperature scenario (Figure \ref{fig:SCM_maxTempEnd}), there is no period during which extreme events happen frequently enough to tip the majority opinion, and so the climate inaction social convention is maintained. The result is emission levels close to the SSP5-8.5 scenario and a much greater increase in temperature than we see in the minimum final temperature case. The results from the largest maximum temperature case (Figure \ref{fig:SCM_maxTempThroughout}) are similar. We note that even in these dire cases there are periods during which the temperature decreases (2030-2060 in Figure \ref{fig:SCM_lateTip}, 2030-2050 in Figure \ref{fig:SCM_maxTempEnd}, and 2095-2100 in Figure \ref{fig:SCM_maxTempThroughout}). These decreases are unrelated to improved climate action, however, since there is not an associated increase in the speaking rate of $A$, $r(t)$. Hence, these temporary decreases in temperature are simply a result of stochasticity in our social-climate model.

%{\sout{We note that even though a year of 12 extreme events occurs in approximately 2060, there are very few extreme events in the following years and these extreme events are not frequent enough to cause a tipping event.}}

\begin{figure}[tbph]
    \centering
    \includegraphics[width=0.3\linewidth]{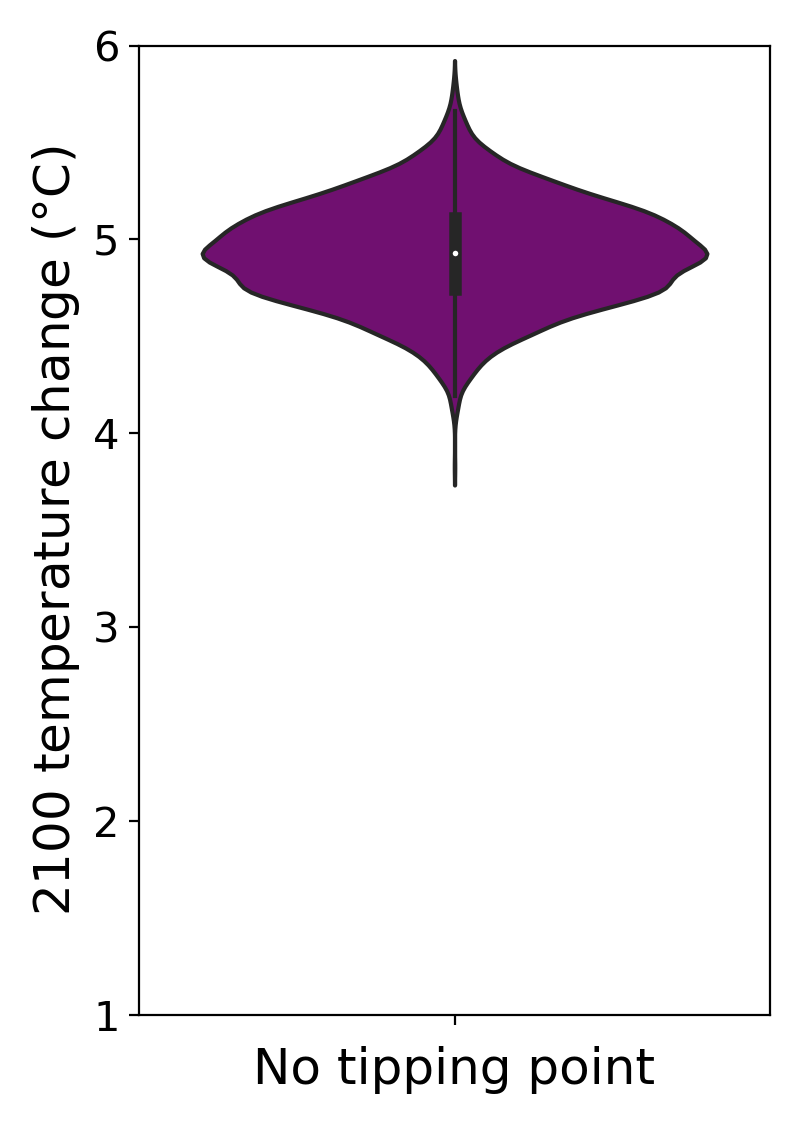}
    \includegraphics[width=0.6\linewidth]{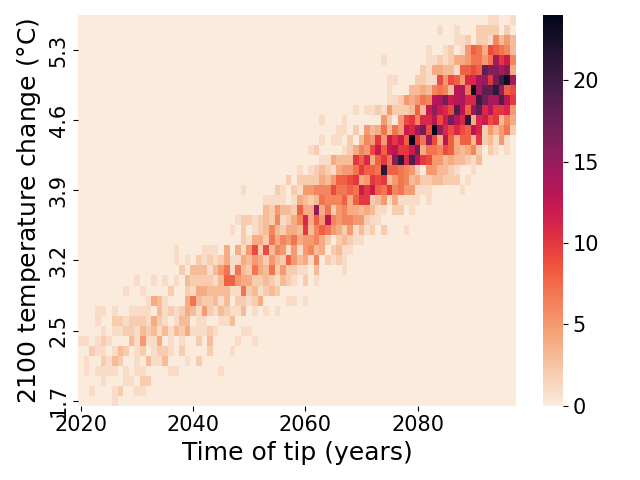}    
    \caption{Left: A violin plot showing the 2100 temperature projections from those simulations wherein the social convention is not overturned. Right: A heat map showing the 2100 temperature projections as a function of the time the social convention is overturned. The colour indicates the number of simulations for which the given result occurs.}
    \label{fig:SCM_tempDisb}
\end{figure}

\begin{figure}[tbph]
    \centering
    \includegraphics[width=0.6\linewidth]{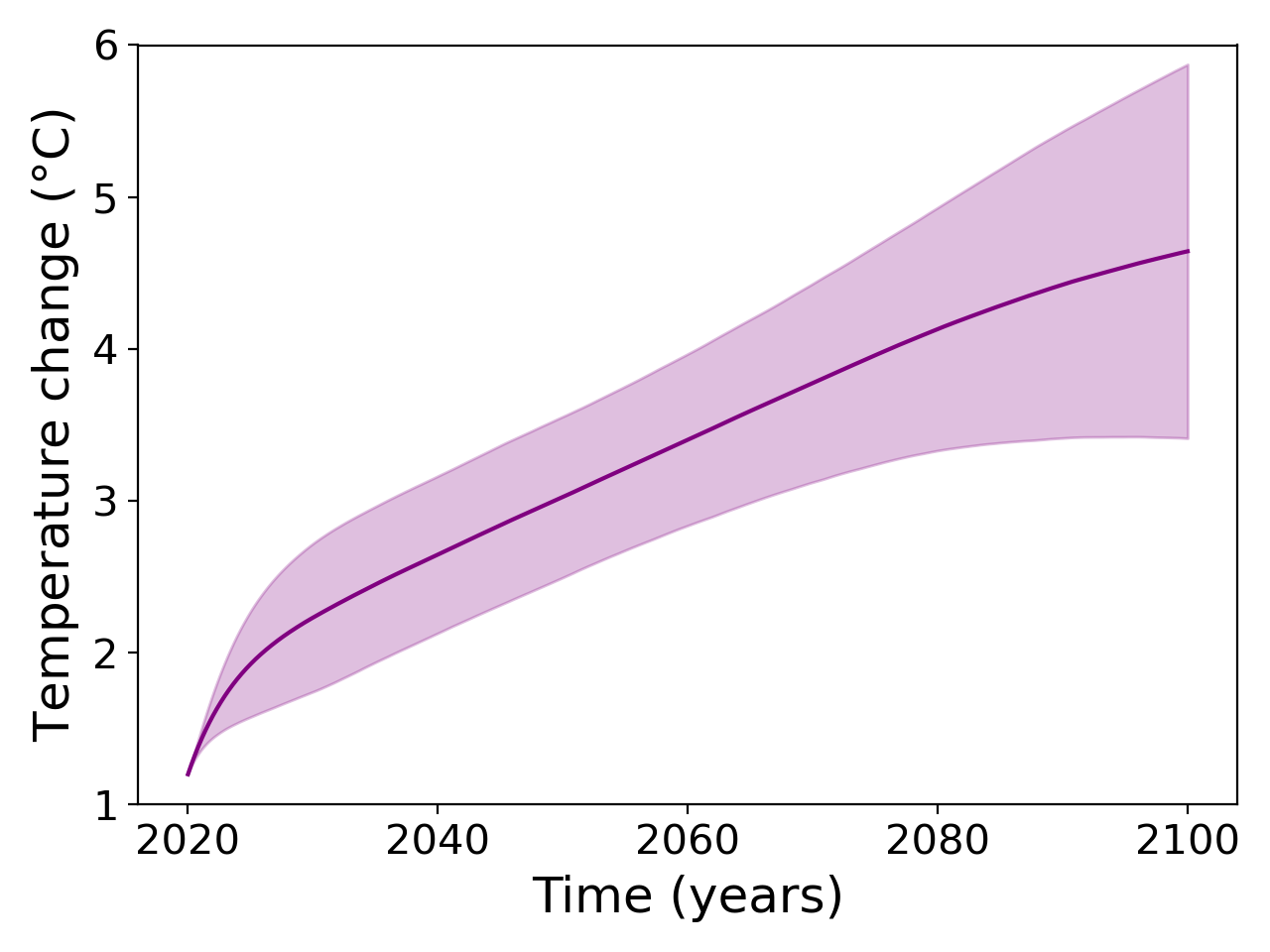}
    \caption{Projections of global surface temperature change from our social-climate model. The solid line shows the average behaviour and the shading indicates the 95\% confidence interval.}
    \label{fig:SCM_avgBehaviour}
\end{figure}

When we consider all of the simulations in which the climate inaction social convention is not overturned (left plot in Figure \ref{fig:SCM_tempDisb}), we find that the social-climate model produces projections similar to the maximum final temperature and largest maximum temperature scenarios, and the median temperature prediction in 2100 is 4.9$\degree$C above pre-industrial levels. These temperature projections are slightly lower than the projections from our SSP5-8.5 climate model (Figure \ref{fig:Climate_SDE}) since the committed minority speaks $A$ and decreases the emissions fed into the climate model. When we consider all of the simulations in which the social convention is overturned (right plot in Figure \ref{fig:SCM_tempDisb}), the final temperature depends on when the tipping point is crossed, i.e., the earlier the tip, the lower the temperature in 2100. We find that about 40\% of our simulations cross the tipping point in the social model and overturn the climate inaction social convention. In general, however, the simulations that do cross the tipping point do so quite late. In these cases, there is not enough time to reverse the effects of increased CO$_2$ emissions, and the final temperatures in 2100 remain similar to those when there is not a tipping event. The average behaviour from our simulations (Figure \ref{fig:SCM_avgBehaviour}) is most similar to that of the SSP3-7.0 climate model.

\section{Discussion} \label{Sect:Discussion}
%------------------------------------------------------------------------------------
% OUR MAIN RESULTS AND GENERAL COMMENTS
%-------------------------------------------------------------------------------------------

%Our social-climate model serves as a framework to expand on current models studying climate change using a dynamic social-climate feedback loop and demonstrates that such a 

Recently, there has been increased interest in human-and-environment models (Complex Human And Natural Systems -- CHANS) that connect environmental processes and human behaviour in a feedback loop, as well as demonstration that this feedback loop has a strong effect on climate outcomes \citep{Beckage2018, Ghidoni2017, Menard2021, Tavoni2011, bechthold:2025, kumar:2025, savitsky:2025}. 
%With increased scientific understanding of climate change, however, there is now evidence that the interaction of the two time scales can be relevant, and does create a feedback loop that can have a strong effect on model outcomes \citep{Beckage2018, Menard2021}. 
Here, we expand on previous work in this area by using a dynamic behaviour model based on fear of extreme events, rather than the game-theoretic or utility function approach that is often used \citep{Tavoni2011, Milinski2008, Menard2021}. Fear can be a strong motivator \citep{Haugestad2025, Lorenzini2023}, and traumatic events such as wildfires and floods, especially if they render people homeless, can have a long-lasting impact on human behaviour \citep{Demski2016, Konisky2015}. Our approach thus allows us to explore the consequences, for climate change, of behaviour that is driven by fear of extreme climatic events, rather than by individuals trying to optimize their own fitness in some way.

Consistent with earlier work, we find that the climate-human behaviour feedback loop may be important in accurately predicting future temperatures \citep{bechthold:2025, kumar:2025, savitsky:2025, Beckage2018, Moore2022, Menard2021}.  Furthermore, we find that the inclusion of dynamic human behaviour increases the variance in climate projections (see Figures \ref{fig:Climate_SDE} and \ref{fig:SCM_avgBehaviour}) and allows us to study how extreme climate events can lead to action altering our future climate. A key result from our work is that the overturning of climate inaction and the timing of this tipping event have a large impact on future temperatures. Specifically, our model confirms that it is important to disrupt the climate inaction social convention and start climate change mitigation strategies sooner rather than later \citep{IPCC_data, Beckage2018, Otto2020, Moore2022}. 

A second key result from our work is that the combination of a committed minority and a sufficient number of extreme events over a short enough time can play a vital role in shifting public opinion towards climate action and causing a social tipping event. It has previously been found that extreme events can shift a population into action \citep{Demski2016, Bergquist2019, Ghidoni2017} and committed minorities often initiate social change and can upset social conventions \citep{Bolderdijk2021}. Indeed, in many locations around the globe the majority opinion in the population may already favour tipping to climate action \citep{smith:2025}, and previous mathematical models find that challenging social conventions can be key to increasing climate action \citep{Beckage2018, Constantino2022, Otto2020}.   Many studies have found a correlation between experiencing extreme events and climate action \citep{Demski2016, Ray2017, Konisky2015} or willingness to act \citep{cologna:2025, gould:2024, Hurlstone2014, Lidskog2015}.  The functional form of this relationship, however, is not known, a gap that hampers prediction efforts. Studies on the effect of extreme events on the population can help define the true relationship between extreme events and human behaviour \citep{Menard2021, Tavoni2011, Demski2016, Bergquist2019, Beckage2018}, in particular, the rate at which the salience of the triggering event decays from memory \citep{Fanta2019, Ray2017, Walshe2020}. \citet{Beckage2018} find that the shape  of this relationship can greatly affect a population's willingness to act at any particular point in time, which then impacts the climate trajectory.

Mathematical climate and social-climate models, including the models in this manuscript, commonly provide projections up to the year 2100 and no further \citep{IPCC_data, Leach2018}. Recent studies suggest that this time span is no longer sufficient to understand the impacts of climate change, and make projections until 2500 or more \citep{Meinshausen2020, Lyon2021, Mengel2018, smith:2025, savitsky:2025}. Indeed, the global average surface temperature is expected to keep increasing past 2100 for all SSP scenarios except SSP1-1.9 and SSP1-2.6 \citep{Lyon2021}. Impacts beyond 2100 include sea level rise even when net-zero GHG emissions are reached by 2050 \citep{Mengel2018}, heat stress that could prove fatal \citep{Lyon2021}, and significant decreases in crop yields \citep{Tigchelaar2018, savitsky:2025}. The CO$_2$ projections up to year 2500 for each of the five SSP scenarios \citep{Meinshausen2020} provide an avenue to expand our current work and make projections on a longer time scale.

% EXTREME EVENTS
%-------------------------------------------------------------------------------------------
\subsection{Extreme Events}

One of our key assumptions is that people identify climate change as the cause of the extreme events they experience and thus respond by advocating for and adopting climate action.  This assumption does have empirical backing \citep{gould:2024, cologna:2025}.  Determining the extent to which climate change is the cause of any given extreme event is a focus of rapid extreme event attribution, which has grown substantially in the last decade \citep{CarbonBrief2022, vanOldenborgh2021, Caldwell2014, Philip2022}. The ``rapid'' component means decreasing the time required to calculate to what extent the event is due to climate change \citep{Oakes2021, Reed2023}. The rapidity with which the results can be published may increase the connection between extreme events and climate change in the general consciousness and thus influence human behaviours \citep{McClure2022, Ogunbode2019, Ghidoni2017}. These attribution studies \citep{CarbonBrief2022, Caldwell2014, vanOldenborgh2021, Philip2022} are also key to fitting parameters such as the base and increased rates of extreme events in our model. We note that our choice of $\lambda_0=4$ as the base rate of extreme events in the absence of climate change is a low estimate. Since we are interested in the qualitative behaviour of this model rather than quantitative predictions, we leave to future work the task of using data from rapid attribution studies to find a better fit for related model parameters. 

By using non-spatial models, we assume that the Earth is a well-mixed system and each uncommitted individual has the same probability of being affected by an extreme weather event. We also assume that each individual changes their behaviour in the same way as a result of experiencing an extreme event. Either or both of these assumptions could be relaxed in future extensions of our work. More accurate  models could include, e.g., spatial effects \citep{Lyon2021, Rantanen2022, Wang2022, Meinshausen2020}, variations in human behavioural responses to extreme events \citep{Tavoni2011, Menard2021, Biswas2009}, or variation in the intensity of extreme events \citep{Mirzaei2025}. The frequency of extreme events is also likely region-specific \citep{Rantanen2022, Mirzaei2025, wang:2013}. %Models that consider spatial effects predict, for example, increased warming of polar regions, a phenomenon called Arctic amplification \citep{Lyon2021, Rantanen2022}. 
%Studies have found that over the last four decades, the Arctic has warmed at least four times faster than the rest of the Earth, with some regions warming seven times faster than the global average \citep{Rantanen2022}.
Models that study variation in  human behaviour find that factors such as financial inequality influence an individual's perception and response to extreme events \citep{Tavoni2011, Menard2021}. Inequality can prevent cooperation \citep{Tavoni2011} and, thus, increase global surface temperatures \citep{Menard2021}. In particular, when the individuals in a model are divided into poor and rich groups, within-group cooperation increases but between-group polarization also increases \citep{Menard2021}. The globally non-uniform distribution of extreme events and economic forces can lead to a similarly non-uniform distribution in climate action, which will likely affect our predictions.  In particular, we expect that non-uniformity will alter the size of committed minority needed to cause a tip toward climate action and the time by which such a tip must occur to prevent extreme warming.

%Possible extensions to our current work include studying group dynamics (e.g., poor versus rich), including economic effects, or using a spatial climate model to account for local variations in extreme events.

% STARTING AND MAINTAINING CLIMATE ACTION
%-------------------------------------------------------------------------------------------
\subsection{Climate Action}
%Extreme events are one of the driving forces to start climate action in our social-climate model. Some scientists conjecture that major disasters are required to motivate climate action \citep{Ghidoni2017} and it has been found that experiencing an extreme event alters an individual's behaviour \citep{Demski2016, Konisky2015}. Additional studies have found that the framing of an extreme event (e.g., whether an extreme event is more or less probable due to human influence) impacts an individual's risk perception \citep{Spence2010} and willingness to act \citep{Lidskog2015, Hurlstone2014}.

%{\sout{In order to increase climate action, we suggest increasing rapid extreme event attribution and clearly stating when extreme events are caused or worsened by climate change.}}

In addition to extreme events, the other driving force for climate action in our model is the effect of the committed minority that holds the climate action opinion. If a committed minority is particularly influential, its opinion is spread through the population faster and more widely than the opinion of a less influential minority \citep{Nyborg2016} and can upset an inaction convention \citep{Beckage2018, Bolderdijk2021, Constantino2022}. In addition, while committed minorities can be a driving force for social change via a bottom-up mechanism, policy makers can contribute by offering top-down support \citep{Bolderdijk2021}. If decision makers do not represent all generations, then resolving issues can be challenging. Studies suggest that decision makers have less incentive to help future generations \citep{Sherstyuk2016} and are more focused on short-term goals \citep{Cseh2018}. An extension to the current work is considering how social interactions between leaders (i.e., highly influential people or decision makers), committed minorities, and citizens (i.e., everyone else) impact climate action and the future climate.

Extreme events and social interactions are not the only driving forces of climate action. It has been found that co-benefits of action can also encourage climate action \citep{Bain2015, Marshall2023}. In particular, \citet{Bain2015} find that care for a greater community, and scientific and economic development, can motivate people to commit to climate action regardless of whether they believe climate change is a risk. The authors suggest that co-benefits of action have a similar impact on human behaviour as the recognition of the importance of mitigating climate change. A similar result is found by \citet{Marshall2023}, who find that people responded most positively to climate action messaging related to helping future generations. \citet{Roggero2023}, however, suggest that the relationship between co-benefits and climate change mitigation is more complex. In particular, the authors found that drivers such as, e.g, international recognition, affected policy changes more than health benefits from decreased air pollution. We leave the study of other driving factors of climate action to future work. 

Many factors, including political ideology, misinformation, and oversimplification, can influence climate change denial and create a committed minority on both sides \citep{McCright2011a, Bliuc2015, McCright2011b}. In particular, \citet{Bliuc2015} note that 97\% of climate science papers hold a consensus on climate change being caused by human behaviour, yet a survey found that only about 50\% of the global population agrees \citep{Kohut2013}. Studies suggest that there is a public divide based on factors such as political ideology, race, and gender \citep{Bliuc2015, McCright2011b}. As a result, part of the population denies climate change, while another part of the population recognizes the need for climate action. Misinformation magnifies this political divide and can lead to climate inaction movements that prevent climate change mitigation policies \citep{Treen2020, Brulle2013}. While work is being done to investigate how to prevent and reverse the spread of misinformation, clear answers do not yet exist \citep{vanderLinden2017, Treen2020}. Extending our social model to include committed minorities holding each opinion would allow us to study the impact of climate change denial. We hypothesize that this extension may result in the social model tipping point becoming reversible. This possible change in model dynamics would also allow for study of ways to maintain climate action once it is initially achieved.

Since it takes a decade to experience the full effects of a GHG emission \citep{Ricke2014}, there is a significant delay between GHG emissions and the associated extreme weather events. When extreme events are delayed, there is a larger increase in cumulative emissions than if extreme events occur immediately after a GHG emission \citep{Ghidoni2017}. As a result of this delay, it may also be difficult to sustain climate action if there is not an immediate improvement in the climate. In fact, it has been found that people tend to prefer policies that produce results in 0-2 years over policies that produce results in 20-30 years \citep{Christensen2020}. It becomes even harder to maintain climate change mitigation efforts when extreme events fade from memory \citep{Walshe2020, Ray2017, Fanta2019}. One model finds that only extreme events that occurred in the previous month have a statistically significant effect on support for climate adaptation policies \citep{Ray2017}. In contrast, another study found that it takes approximately two generations for a population to move back into an area previously devastated by a catastrophic flood \citep{Fanta2019}. If delay dynamics were included in our social-climate model or if the social tipping point were reversible (i.e., if individuals commitment to climate action could wane), we might expect oscillatory dynamics where the tipping point is crossed and reversed multiple times \citep{vanderBolt2018}. The social effects of such dynamics are likely to be highly non-trivial.

\section*{Funding}
This work was supported by Natural Sciences and Engineering Research Council of Canada (grant numbers RGPIN-2016-0577 (RCT), RGPIN-2024-04653 (EF), CGS-M (SKW)). 

\section*{Competing Interests}
The funding sources were not involved in the conduction of research or preparation of this article.  The authors have no relevant financial or non-financial interests to disclose.

\section*{Author Contributions}
This work forms part of SKW's MSc thesis.  SKW led the development of the model, wrote the code, ran the simulations, and wrote the first draft of the paper.  EF and RCT supervised SKW's work, providing support with mathematical, coding, and interpretation issues when they arose.  EF and RCT provided significant editorial feedback and support in the process of finalising the paper.

\section*{Data Availability}
The associated code is available at \url{https://github.com/sarahwyse/SocialClimateModel.git} and the simulations in this manuscript are run in Python 3.9.12 (Spyder 5.4.3). 

\printbibliography
% \bibliographystyle{spmpscinat.bst}
% \bibliography{Sources.bib}

\end{document}